\documentclass[preprint1,floatfix,twocolumn ]{emulateapj}
\usepackage{apjfonts}

\usepackage{amsmath, amsbsy}
\usepackage{graphicx}
\usepackage{natbib, hyperref}

\shorttitle{Probing the Expansion history of the Universe}
\shortauthors{Feng \& Li}

\begin{document}

\title{Probing the Expansion history of the Universe  by Model-Independent Reconstruction \\
from Supernovae and Gamma-Ray Bursts  Measurements}

\author{Chao-Jun Feng, Xin-Zhou Li}
\affiliation{Shanghai United Center for Astrophysics (SUCA), \\ Shanghai Normal University,
    100 Guilin Road, Shanghai 200234, P.R.China \\
    Email: fengcj@shnu.edu.cn, kychz@shnu.edu.cn}


\begin{abstract}
To probe the late evolution history of the Universe, we adopt two kinds of optimal basis systems. One of them is constructed  by performing the principle component analysis (PCA) and the other is build by taking the multidimensional scaling (MDS) approach.  Cosmological observables such as the luminosity distance can be decomposed into these basis systems. These basis are optimized for different kinds of cosmological models that based on different physical assumptions, even for a mixture model of them.  Therefore,  the so-called feature space that projected from the basis systems is cosmological model independent, and it provide a parameterization for  studying and reconstructing  the Hubble expansion rate from the supernova luminosity distance and even gamma-ray bursts (GRBs) data with self-calibration. The circular problem when using GRBs as cosmological candles is naturally eliminated in this procedure. By using the  Levenberg-Marquardt (LM) technique and the Markov Chain Monte Carlo (MCMC) method, we perform an observational constraint on this kind of parameterization. The data we used include  the  "joint light-curve analysis" (JLA)  data set that consists of $740$ Type Ia supernovae (SNIa) as well as $109$ long gamma-ray bursts with the well-known Amati relation.
\end{abstract}


\maketitle


\section{Introduction}\label{sec:introduction}
One of the major target for present observations is to learn the evolution history of the Universe through the cosmic expansion rate. The observations of the  Type Ia supernovae (SNeIa) have indicated that the Universe is currently accelerating \citep{Riess1998,Perlmutter1999}.  For the lack of deeper understanding, the cause of this acceleration is  usually explained by introducing an exotic energy component called dark energy. There are many dark energy models that based on different physical origins, see \citet{Li2011} for a recent review.  A specific dark energy model is usually  characterized by a small set of parameters. One can constrain these parameters by observational data to obtain the expansion rate of the Universe. Although this approach is reasonable, the result is often depending on which model one used. So an interesting  question is that how to probe the  cosmic evolution history from observations without any reference to a specific dark energy model.

All such researches are often called the cosmological model-independent  reconstruction of the cosmic expansion rate from observations, and it has been largely discussed in the literature. Most of them are based on a smoothing procedure in redshift bins \citep{Huterer2003}.  \citet{Crittenden2005, Simpson2006}  have also performed  PCA to  reconstruct the dark energy equation of state. \citet{Mignone2008} has expanded the luminosity distance into a series of orthonormal functions as basis to reconstruct the cosmic expansion rate. \citet{Maturi2009} has optimized this basis system to be capable to describe cosmologies independently of their background physics. The quality of the estimation of the luminosity distance is also improved. \citet{Li2014} has applied this method to determine the curvature parameter. 

Gamma-Ray Bursts (GRBs) are the most intense explosions in the Universe, and they can potentially be another standard candles living in the high redshifts.  There are many GRBs observed at $0.1 < z \leq 8.1$, whereas the maximum redshift could to be $10$ or even larger in the future observations. So, GRBs is a complementary probe to SNeIa, see \citet{Schaefer2007} for a review on the so-called GRB cosmology. However, there is a circularity problem when using GRBs as cosmological candles, because low-redshift GRBs at $z<0.1$ are too few to calibrate the correction relation in a model-independent way. Then, an input cosmology is needed to obtain the relation, but it leads to the circular problem when constraining cosmological parameters. To alleviate the circularity problem, some statistical methods have been proposed in \citet{Ghirlanda2004}, such as the scatter method, the luminosity distance method, and the Bayesian method in \citet{Firmani2005}. \citet{Liang2008, Kodama2008} have suggested calibrating GRBs by using the SNeIa data, see also \citet{Wei2010} for a relevant work. Another interesting approach was proposed by  \citet{Li2008}, in which they have treated the parameters involved in GRBs as free parameters and determined them simultaneously with other cosmological parameters by global fitting.

In this paper, we adopt two kinds of optimal basis systems to probe the the evolution history of the Universe. One of them is constructed  by performing the principle component analysis (PCA) following the way of \citet{Mignone2008, Maturi2009}. But there are some differences, which will be discussed in the next section.The other  kind of optimal basis is build by taking the multidimensional scaling (MDS) approach \citep[chapter 5]{Borg2013}, which is another powerful method to reconstruct the cosmic expansion rate. These basis have been optimized for different kinds of cosmological models that based on different physical assumptions, even for a mixture model of them. Therefore,  the so-called feature space that projected from the basis systems is cosmological model independent, and it provide a parameterization for  studying and reconstructing  the Hubble expansion rate from the supernova luminosity distance and even gamma-ray bursts (GRBs) data with self-calibration. By using the  Levenberg-Marquardt (LM) technique and the Markov Chain Monte Carlo (MCMC) method, we perform an observational constraint on this kind of parameterization. The data we used include  the  "joint light-curve analysis" (JLA)  data set that consists of $740$ Type Ia supernovae (SNIa) as well as $109$ long GRBs with the well-known Amati relation. The circular problem when using GRBs as cosmological candles is naturally eliminated in this procedure. This may look like the global fitting method proposed by \citet{Li2008},  but here we do not assume any cosmological models in advance. 

The structure of this paper is as follows. In Section \ref{sec:method}, we present the essential parts of the model-independent method and show how efficient when these methods  are applied to optimize the basis for different kinds of cosmological models. The description of data and application of the method to reconstruct the evolution of the Universe are shown in Section \ref{sec:application}. The discussions conclusions are presented in Section \ref{sec:discon}.

\section{Model-Independent Method}\label{sec:method}
In this section, we will present applications of PCA and MDS to the JLA and GRB data to probe the evolution of the Universe.  At first, we will give the basic formulae and expand cosmological observables into a finite sums of functions as basis. Then, the basis are optimized by using the PCA and MDS methods respectively.  In fact, the basic ideas of PCA and MDS are very similar and one can finally obtain the most important components that could be used to  describe the observables.  It should be noticed that we will directly focus on the  cosmological observables like distances instead of the physical quantities within a specific cosmological model , such as  the equation of state of  dark energy.

\subsection{Basic formulea for the cosmic expansion}

In  Friedmann-Robertson-Walker metric, the luminosity distance is given by
\begin{equation}\label{equ:lumdis}
	D_L(z) = \frac{c}{H_0} \frac{1+z}{\sqrt{|\Omega_k}|} \text{sinn} \left( \sqrt{|\Omega_k|} \int_0^z  \frac{dz'}{E(z')}\right) \,,
\end{equation}
with $E(z) = H(z)/H_0$, and sinn$(x) = \sin(x), x, \sinh(x)$ for $k=1,0,-1$ respectively. Here, $c$ is the speed of light, and $\Omega_k \equiv - k/(a_0H_0)^2$ denotes the density of the spatial curvature at present. By taking the derivative of Eq.(\ref{equ:lumdis}) with respect to the redshift $z$, one can obtain
\begin{equation}\label{equ:history}
	E(z)^{-1} = \frac{D'(z)}{\sqrt{1 + \Omega_k D^2(z) }} \,,
\end{equation}
where $D(z)$ is the $H_0$-independent comoving angular diameter distance that relates to the luminosity distance as
\begin{equation}
	D(z) = \frac{H_0}{c} \frac{D_L(z)}{ (1+z)} \,,
\end{equation}
and where the prime  denotes the derivative with respect to the redshift $z$.
For a flat universe, Eq.(\ref{equ:history}) could be written as
\begin{equation}\label{equ:history2}
	E(z)^{-1} =  \frac{H_0}{c}  \left[ \frac{D_L'(z)}{1+z} - \frac{D_L(z)}{(1+z)^2}\right] \,,
\end{equation}
Obviously, if the behaviours of both $D(z)$ (or $D_L$) and its derivative could be dug up from some observational data, one can obtain the evolution history of the universe from Eq.(\ref{equ:history}) or (\ref{equ:history2}). Although the data from observations of SNe Ia provide measurements of the distance  modulus and redshifts, it is not a convenient way to taking derivative to the luminosity distance directly from the data, because  the result would be extremely noisy and unreliable.  Therefore, we need to first properly smooth the data by fitting an adequate function $D(z)$ to the measurements in a model-independent way. The derivative can then be approximated by the derivative of  $D(z)$.  This can be achieved through an expansion of $D(z)$ into a finite sums of suitable functions $p_i(z)$ like:
\begin{equation}\label{equ:expan1}
	D(z) = \sum_{i=1}^{M} c_i p_i(z) \,.
\end{equation}
The $M$ coefficients $c_i$ can be determined by fitting the data, namely $c_i$ are those which minimize the $\chi^2$ statistic function. The number of the terms to be included in the expansion depends on the choice of the orthonormal basis and the quality of the data.  The basis $\{p_i\}$ could be arbitrary with idea data, but it will bot be in practice.  \citet{Benitez2012} has used the Gram-Schmidt orthonormalization to decompose the luminosity distance, and they found a systematic trend on the slope of the reconstructed cosmic expansion rate. It indicated that a randomly chosen system of orthonormal basis functions may not be well adapted to the behavior of the measured data. \citet{Maturi2009}  has suggested optimizing the basis system by using of PCA to reduce the number of coefficients $M$ in Eq.(\ref{equ:expan1}) , and the possible bias introduced by the choice of the basis is also removed, see \citet{Benitez2013}. In this paper, we will make use  of two optimal basis systems that one derived from the principal component analysis (PCA) and the other one from the multidimensional scaling (MDS) approach. The number of coefficients required  is minimized by either PCA or MDS methods. Besides, it also removes any bias introduced by the choice of the basis.

 \subsection{The Optimal Basis}
 \subsubsection{The Training Set and its Generator}

 To obtain the optimal basis, we start by writing  the $H_0$-independent comoving angular diameter distance in a column vector $\mathbf{ D} = (D(z_1), D(z_2), \cdots, D(z_n))^{T}\in \mathbb{R}^{n}$, which can be regarded as a single point in an $n$-dimensional space.  In the literature,  $n$ is often taken to be the number of data points from observations, but we will expand the variables at the redshifts in a certain range with  a small interval, say $0.1$. And then we apply the  spline  interpolation method to calculate the distance at data points. Also, the range of the redshifts is enlarged to cover that of the GRBs. 

 Now we select a group of  models that are believed to space the set of variable cosmologies and calculate $\mathbf{ D}$ for each model to generate a set of vectors $\mathbf{ D}_i $  with $i=1,2,\cdots, M$, where $M$ is the number of models. The ensemble of models $\mathbf{T} = (\mathbf{D}_1 , \mathbf{D}_2 , \cdots, \mathbf{D}_M ) \in \mathbb{R}^{n\times M} $ are called the  \textit{training set}  introduced by \citet{Maturi2009}.  In principle, the train set could be constructed from any models with arbitrary functions, but it is convenient to consider the models at least weakly resembling the data set \citep{Maturi2009}. In other words, one can choose any models, as long as the data set is tightly enclosed in the distribution of the $M$-point cluster in the $n$-dimensional space \citep{Maturi2009, Benitez2013, Li2014}.  To avoid confusion with a specific cosmological model that determines the evolution of the universe, we would like to call these models the \textit{training set generators} (TSGs), which mean they are only responsible for building the training set. 

In the literature, the $\Lambda$CDM model with parameters uniformly sampled in the parameter space are often considered as a TSG to build the training set, but of course other kinds of cosmological models can be used as well, such as the dynamical dark energy models, modified gravity models, or even a mixture of them. However, the result optimal basis system is independent of any TSGs.  To see this, we will take the non-flat $\Lambda$CDM,  the wCDM, the Chevallier-Polarski-Linder (CPL) parametrization model \citep{Chevallier2001, Linder2003}, the FSLL parametrization model without divergence \citep{Feng2012}, the holographic dark energy (HDE) model \citep{Li2004}, the Dvali-Gabadadze-Porrati (DGP) model \citep{Dvali2000, Deffayet2001,Deffayet2002},  the new agegraphic dark energy (NADE) model \citep{Wei2008},  the Ricci dark energy (RDE)  model \citep{Gao2009, Feng2009} and their mixture as the TSGs. It also shows that no matter which kind of TSG we used, the dimensionality of the training set could be reduced efficiently by using either the principal component analysis~(PCA) or the multidimensional scaling analysis~(MDS).

\subsubsection{Building the Optimal Basis with PCA}
PCA is a very useful statistical tool to reduce the dimensionality of an initially large training set space.  Taking the mean of the training set, we obtain a reference model $\mathbf{D}_{\text{ref}}$ that defines the origin of the $n$-dimensional space:
\begin{equation}\label{equ:refmod}
	\mathbf{D}_{\text{ref}} = \langle \mathbf{D}_i \rangle = \frac{1}{M} \sum_{i = 1}^{M} \mathbf{D}_i \, \in  \mathbb{R}^{n\times 1} \,.
\end{equation}
Then, one can define  the so-called \textit{covariance matrix}  by :
\begin{equation}\label{equ:cov}
	\mathbf{S} = \frac{1}{M}\boldsymbol{\Delta \Delta^{T} } \,,\quad \text{with} \quad \boldsymbol{\Delta} = \mathbf{T} - \mathbf{D}_{\text{ref}} A \, \in  \mathbb{R}^{n\times M} \,,
\end{equation}
where $A = (1, 1, 1, \cdots, 1) \in  \mathbb{R}^{1\times M} $.  Therefore, the principle components (PCs) are the eigenvectors of the $\mathbf{S}$ matrix, which can be obtained by solving the eigenvalue problem $\mathbf{S w}_i = \lambda_i  \mathbf{w}_i$.  In the following, the eigenvalues $\lambda_i (i=1,2,\cdots, n)$ are sorted in a descendent sequence $\lambda_i > \lambda_{i+1}$,  and the corresponding eigenvectors $\mathbf{w}_i$  are called the first PC ($\mathbf{w}_1$), the second PC ($\mathbf{w}_2$),  and so on.   This gives us  the components in order of significance, and we can decide to ignore the components of lesser significance. For instance, if we choose only the first $p$ eigenvectors, then the information content of the training set can be optimised via a linear transformation $\mathbf{W}$:  $\mathbb{R}^{n} \rightarrow \mathbb{R}^{p}$ mapping the training set vectors into a so called \textit{feature space}: $\mathbf{t}_i= \mathbf{W^T D}_i\in \mathbb{R}^{p}, (i = 1, 2, \cdots, M)$.  Here, $\mathbf{t}_i$ are called the \textit{feature vectors}, while the linear transformation is given by: $\mathbf{W}= (\mathbf{w}_1, \mathbf{w}_2, \cdots,\mathbf{w}_p) $.  We do lose some information for ignoring $(\mathbf{w}_{p+1}, \mathbf{w}_{p+2}, \cdots, \mathbf{w}_n)$, but if their eigenvalues are small enough, we do not lose much.  Then, one could expand $D(z)$ into the optimal basis as
\begin{equation}\label{equ:exppca}
	D(z) = \mathbf{D}_{\text{ref}} + \sum_{i=1}^p c_i \mathbf{w}_i \,,
\end{equation}
with some coefficients $c_i$ that will be determined by fitting data through $\chi^2$ minimization.  The $D'(z)$ is derived by taking derivative with respect to the redshift on both side of Eq.(\ref{equ:exppca}).  Since the eigenvalue $\lambda_i $ is just the variance of $\Delta$ along the vector $w_i$,  the percentage of variance we are willing to consider will then determine the number of PCs to be included in the reconstruction matrix $\mathbf{W}$, i.e. the value of $p$. For example, we define the cumulative percentage of total variation \citep[section 6.1.1]{Benitez2013, Jolliffe2002} as:
\begin{equation}\label{equ:threpca}
	 r_p = \frac{\sum_{i=1}^p \lambda_i }{\sum_{i=1}^n \lambda_i}  \,,
\end{equation}
and after setting a threshold, e.g. $r_p>99\%$,  it will return the value of $p$.

\subsubsection{Building the Optimal Basis with MDS }
MDS is another useful statistical tools to reduce the dimensionality of the training set.  There are many types of MDS \citep[chapter 5]{Borg2013}, which can be classified according to whether the similarities data are qualitative (called non-metric  MDS) or quantitative (called metric MDS). In this paper, we will take the algorithms of so-called  the classical MDS (CMDS) , a special kind of metric MDS. In CMDS, a single  Euclidean distance matrix is often used. From the training set $\mathbf{T}$ built before, one can easily construct a square-distance matrix  $\mathbf{Q}$ \citep[chapter 5]{Borg2013}, whose components are given by 
\begin{equation}\label{equ:distance}
	Q_{ij}= \sum_{k=1}^n (D_{ik} - D_{jk})^2 \, \in \mathbb{R}^{M\times M}\,,
\end{equation}
with $i=1,2,\cdots, M$. The matrix $\mathbf{Q}$ describes the dissimilarity of a pair of $\mathbf{D}$s. Centering the matrix $\mathbf{Q}$, we obtain the  Gram matrix of $\mathbf{Q}$:
\begin{equation}\label{equ:gram}
	\mathbf{G} = -\frac{1}{2} \mathbf{Z Q Z} \,,
\end{equation}
 where $\mathbf{Z=I_M - M^{-1} 11^T}$ with $I_M$  the identity matrix of order $M$, and $\mathbf{1}$ a vector with a $1$ in each of its entries.  Then, we compute the eigenvalues and eigenvectors of the matrix $\mathbf{G}$, $\lambda_i$, $y_i$. And as before, $\lambda_i$ are sorted in a descendent sequences $\lambda_i > \lambda_{i+1}$. Therefore, by taking the first $p$ positive eigenvalues  and the corresponding first $p$ eigenvectors,  we get the MDS configuration with low dimension $p<M$ as  $\mathbf{X} = \mathbf{Y_+} \Lambda_+^{1/2}\, \in \mathbb{R}^{M\times p} $, where $\mathbf{Y}_+ = (\mathbf{y}_1,\mathbf{y}_2,\cdots,\mathbf{y}_p)$ and $\Lambda_+ = \text{diag}(\lambda_1, \lambda_2, \cdots, \lambda_p)$. Here, $\mathbf{X^T}\in \mathbb{R}^{p\times M}$ plays the same role of the feature space in the PCA mapping, i.e. $\mathbf{t}= \mathbf{\tilde W^T T} \in \mathbb{R}^{p\times M} $. Finally, one could expand $D(z)$ into
\begin{equation}\label{equ:expmds}
	D(z) = \sum_{i=1}^p \tilde{c}_i \mathbf{\tilde w}_i\,,
\end{equation}
where the optimal basis are given by
\begin{equation}\label{equ:mdsbasis}
 	\mathbf{\tilde W}= (\mathbf{\tilde w}_1, \mathbf{\tilde w}_2, \cdots,\mathbf{\tilde w}_p)
	= \mathbf{T T^{T}}  \mathbf{T Y_+\Lambda}^{1/2}_+\Sigma \, \in  \mathbb{R}^{n\times p}\,.
\end{equation}
Here $\Sigma $ is a diagonal matrix to rescale the basis, such that  the maximum absolute value of each $\mathbf{\tilde w_i}$ is equal to one. We shall find that this model is also well consistent with observations.   Defining the following two cumulative quantities
\begin{equation}\label{equ:thremds}
	r_{p}^{(1)} = \frac{\sum_{i=1}^p \lambda_i }{\sum_{i=1}^n |\lambda_i |}  \,, \quad r_{p}^{(2)} = \frac{\sum_{i=1}^p \lambda_i ^2}{\sum_{i=1}^n \lambda_i ^2}  \,,
\end{equation}
we can determine the value of $p$  by either of the thresholds is satisfied, e.g. $r_{p}^{(1)} > 99\%$, or $r_{p}^{(2)} > 99\% $.

\subsection{Efficiency of  Different TSGs}

For each TSG mentioned before, we  construct the training set for $20$ times and average the values of $r_1$ and $r_2$ for the PCA method, and the values of $r_1^{(1,2)}$ and $r_2^{(1,2)}$ for the MDS method.  In each training set, there are $20$ models except for the mixture one. In each model,  the range of the redshift is $z\in[0,10)$ with an interval of $0.1$, and the parameters of the TSGs are uniformly sampled with boundaries listed as the last column of Tab.\ref{tab:eff} to calculate the distance $\mathbf{D_i}, (i=1,2,\cdots, 20)$, so that
the training set $\mathbf{T}  \in  \mathbb{R}^{100\times 20}$.  Results are summarized in Tab.\ref{tab:eff}. It should be noticed that the training set from the mixture TSG contains $100$ models that randomly chosen from the other TSGs in Tab.\ref{tab:eff} with the same parameters' ranges. From Tab.\ref{tab:eff}, one can see that the value of $p$  satisfing Eq.(\ref{equ:threpca}) or (\ref{equ:thremds}) could be very small, say, $p=1, 2$, whenever which TSG is used.

\begin{figure}
  \epsscale{1.0}
   \plotone{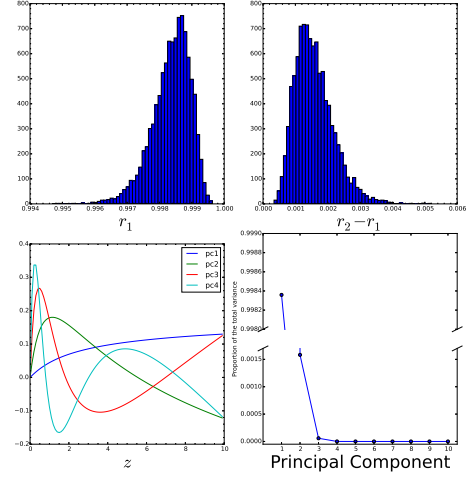}
  \caption{ PCA method. Top: Histograms of the first PC (left) and second PC (right) in the percentage of total variation from the $\Lambda CDM$ TSG. Bottom-left: The first $4$ PCs for the comoving angular diameter distances. Bottom-right: The scree plot. All the values of parameters are uniformly sampled with boundaries $0.1<\Omega_m<0.9, 0.1<\Omega_{\Lambda}<0.9$ and $-0.1<\Omega_k<0.1$  as listed in the Tab.\ref{tab:eff}.  The range of the redshift is $z\in [0, 10)$ with an interval of $0.1$. }
  \label{fig:PCA}
\end{figure}

\begin{figure}
  \epsscale{1.0 }
   \plotone{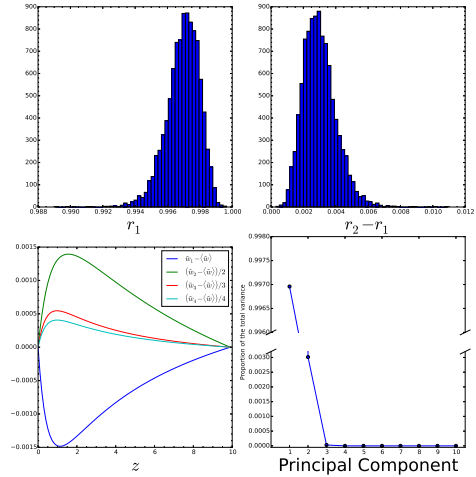}
  \caption{MDS method. Top: Histograms of the first eigenvalue (left) and second eigenvalue (right) in the percentage of total absolute eigenvalues from the $\Lambda CDM$ TSG. Bottom-left: The first $4$ basis for the comoving angular diameter distances. To get a better visualization to see how they differ one from the other, we plot their difference with their average and then divided by their number: $(\tilde w_i - \langle \tilde w\rangle)/i$ for $i=1,2,3,4$. Bottom-right: The scree plot. All the values of parameters are uniformly sampled with boundaries $0.1<\Omega_m<0.9, 0.1<\Omega_{\Lambda}<0.9$ and $-0.1<\Omega_k<0.1$  as listed in the Tab.\ref{tab:eff}.  The range of the redshift is $z\in [0, 10)$ with an interval of $0.1$. }
  \label{fig:MDS}
\end{figure}

We have chosen the  $\Lambda CDM$ model as the TSG to build the training set for $10000$ times and plotted the distribution of the values of $r_{1}$ in the top-left panel of Fig.\ref{fig:PCA} for the PCA analysis, and for each time the number of models is uniformly sampled from $20$ to $50$.  It shows that the first principal component (PC) has the largest possible variance, namely that it retains $>99.0\%$ of the total variance in the sample. For comparison,  the histogram of the second PC in the percentage of total variation, i.e. $r_2 - r_1$  is also plotted in the top-right panel of Fig.\ref{fig:PCA}.  Therefore, the first two PCs retain $>99.99\%$ of the total variance,  which means that they have already considered the major properties in the expansion of the training set.  For the MDS approach, we get almost the same results when using the first kind of threshold in Eq.(\ref{equ:thremds}),  see the two top panels of Fig.\ref{fig:MDS}.  However, when the second kind of threshold in Eq.(\ref{equ:thremds}) is applied, $r_1$ has already preserved $>99.99\%$ of the total squared of the eigenvalues, see Tab.\ref{tab:eff}.

The bottom-left panel of Fig.\ref{fig:PCA}  depicts the first four PCs for the comoving angular diameter distances, while the bottom-right one shows the scree plot. It is clear that the feature space with  $2$-dimension are enough to describe the distance vector without losing much information. The same conclusions could be drawn from  Fig.\ref{fig:MDS}. In fact, by using the MDS approach,  the the feature space with  $1$-dimension is good enough.

\section{The Cosmic Expansion History  Reconstruction}\label {sec:application}
Since the feature spaces discussed before retain all significant cosmological information, they can be used to parameterize cosmologies,  \citet{Maturi2009} called the principal components  cosmological eigen-modes  (\textit{eigen-cosmologies}). They aim to describe observable quantities directly, while the "standard" cosmological parameters describe the physical properties.  To  make a distinction between the parameterization from PCA method Eq.(\ref{equ:exppca}) and that from MDS method Eq.(\ref{equ:expmds}), we will call them the \textit{PCA-model} and the \textit{MDS-model} respectively.  In the following, these two models are fitted by the SNeIa and GRBs data.  Then, the evolution history of the Universe is obtained by using these two models.

\subsection{Data Descriptions}

\subsubsection{JLA Supernovae Data}
The latest large SNeIa data set  is the  "joint light-curve analysis" (JLA)  sample, in which it contains $740$  spectroscopically confirmed type Ia supernovae covering the redshift range $0.01<z<1.3$ with high quality light curves. The distance estimator in this analysis assumes hat supernovae with identical color, shape and galactic environment have on average the same intrinsic luminosity for all redshifts. This hypothesis is quantified by a linear model, yielding a standardized distance modulus \citep{Betoule2014, Shafer2015}
\begin{equation}\label{equ:modulusobs}
	\mu_{\text{obs}} = m_{\text{B}} - (M_{\text{B}} - A \cdot s + B \cdot C + P \cdot \Delta_M)  \,,
\end{equation}
where $m_{\text{B}}$ is the observed peak magnitude in rest-frame B band, $M_{\text{B}}, s, C$ are the absolute magnitude, stretch and color measures, which are specific to the light-curve fitter employed, and $P(M_* >10^{10} M_\odot)$ is the probability that the supernova occurred in a high-stellar-mass host galaxy. The  stretch, color, and host-mass coefficients ($A, B, \Delta_M$, respectively) are nuisance parameters that should be constrained along with other cosmological parameters. On the other hand, the distance modulus predicted from a cosmological model  for a supernova at redshift $z$ is given by
\begin{equation}\label{equ:modulusmod}
	\mu_{\text{model}} (z, \vec\theta)= 5\log_{10} \left[ \frac{D_L(z)}{10\text{pc}}\right]\,,
\end{equation}
where $\vec\theta$ are the cosmological parameters in the model, and $D_L(z)$ is the luminosity distance. For a given pair of the heliocentric-frame and the CMB-frame redshifts $(z_{hel}, z_{cmb})$ from the JLA data, 
\begin{eqnarray}
\nonumber
	D_L(z=z_{cmb}) &=& \frac{c}{H_0} \frac{1+z_{hel}}{\sqrt{|\Omega_k}|} \text{sinn} \left( \sqrt{|\Omega_k|} \int_0^{z_{cmb}}  \frac{dz'}{E(z')}\right)\\
	&=&(1+z_{hel})r_A(z_{cmb}) \,, \label{equ:lumdis2}
\end{eqnarray}
where $r_A(z)$ is the comoving angular diameter distance. The $\chi^2$ statistic is then calculated in the usual way
\begin{equation}\label{equ:chi2sn}
	\chi^2_{\text{SN} } = (\vec \mu_{\text{obs}}  - \vec \mu_{\text{model}})^T \mathbf{C_{\text{SN}}}^{-1}  (\vec \mu_{\text{obs}}  - \vec \mu_{\text{model}})\,,
\end{equation}
with $\mathbf{C_{\text{SN}}}$ the covariance matrix of $\vec \mu_{\text{obs}}$.

\subsubsection{GRBs data}

The GRBs data we will use is compiled by \citet{Amati2000, Amati2008, Amati2009}, in which there are $109$ long GRBs with measured redshift ($0.1<z\leq 8.1$) and spectral peak energy. There  are $50$ GRBs at $z<1.4$, and $59$ GRBs at $z>1.4$ in this data set, see Ref.\citep[Table I, II]{Wei2010}. The well-known Amati correlation \citep{Amati2002} in GRBs is given by
\begin{equation}\label{equ:amati}
	\log_{10}\frac{E_{\text{iso}} }{1\text{erg} }=  \lambda + b \log_{10}\left(\frac{E_{\text{p,i}}}{\text{300KeV}} \right)
\end{equation}
where $E_{\text{iso}}$ is the  isotropic-equivalent radiated energy, while $E_{\text{p,i}}$  is the cosmological rest-frame spectral peak energy. Here, $\lambda$ and $b$ are constants to be determined by observations, see \citet{Wei2008}. The isotropic-equivalent radiated energy $E_{\text{iso}}$ is related to the bolometric fluence $S_{\text{bolo}}$ of gamma rays in the GRB at redshift $z$:
\begin{equation}\label{equ:sbolo}
	E_{\text{iso}} = 4\pi D_{L}^2 S_{\text{bolo}} (1+z)^{-1} \,.
\end{equation}
Then, from the GRBs data one can obtain the distance modulus as:
\begin{equation}\label{equ:mug}
	\mu_{\text{g}} = \frac{5}{2}\log_{10} \bigg[  \frac{(1+z)}{4\pi} \left(\frac{E_{\text{p,i}}}{\text{300KeV}} \right)^b \frac{ S_{\text{bolo}}^{-1} }{ 100\text{pc}^2}\bigg]  + \frac{5\lambda}{2}\,,
\end{equation}
with uncertainties
\begin{equation}\label{equ:muerrg}
	\sigma_{\mu_\text{g}}^2 = \left(\frac{5}{2 \ln 10}\right)^2  \bigg[  b^2\left(\frac{\sigma_{E_{\text{p,i}}}}{E_{\text{p,i}}} \right)^2
	+  \left( \frac{\sigma_{S_{\text{bolo}}}}{S_{\text{bolo}}} \right)^2 +  \sigma_{\text{sys}}^2 \bigg]  \,.
\end{equation}
The $\chi^2$ statistic is then calculated by
\begin{equation}\label{equ:chig}
	\chi^2_{\text{g}} = \sum _{i=1}^N \frac{(\mu_{\text{g}} - \mu_{\text{model}} )^2}{\sigma_{\mu_{\text{g}}}^2} \,,
\end{equation}
with $N$ data points. Here $ \sigma_{\text{sys}}$ in Eq.(\ref{equ:muerrg}) denotes the systematic error, which accounts the extra scatter of the luminosity relation.

In literature, the value of  $ \sigma_{\text{sys}}$  is often estimated by finding the value such that a $\chi_g^2$ fit to the luminosity calibration curve produces a reduced $\chi_g^2$ of unity, see Ref.\citep{Schaefer2007}.  In fact, the systematic error should not depend on the number of data points $N$. Based on this assumption, we randomly choose  a subset of the whole $109$ GRBs data set, i.e. $N=20,30, \cdots 100$.  Then, we find the value of  $ \sigma_{\text{sys}}$ such that the reduced $\chi_g^2$ is unity. We have performed this procedure for $100$ times and averaged the value of $ \sigma_{\text{sys}}$, then  presented them in Tab.\ref{tab:seer}.  Also, the standard deviations of  $ \sigma_{\text{sys}}$ is given in  the same table. Finally, we obtained averaging systematic error (weighted by the stander deviations) as 
\begin{equation}\label{equ:syserr}
	 \sigma_{\text{sys}} = 0.7571 \,,
\end{equation}
which will be used in the next fitting procedures. Besides, from Tab.\ref{tab:seer}, it is clear that the $\sigma_{\text{sys}}$ depends on the model through the $\chi^2_g$.

 \begin{deluxetable}{cllll}
\tablecaption{System error evaluations with its standard deviations\label{tab:seer}}
\tablewidth{0pt}
\tablehead{
\colhead{\# of data} & \multicolumn{2}{c}{PCA} &  \multicolumn{2}{c}{MDS} \\
\colhead{}
& \colhead{ $\langle \sigma_{\text{sys}} \rangle$}           & \colhead{ std.}
& \colhead{ $\langle \sigma_{\text{sys}} \rangle$}           & \colhead{ std.}
}
\startdata
$20$	& $0.7910$ &	$0.1361$ &	 $0.7637$ & $0.1321$ \\
$30$	& $0.7445$ &	$0.0828$ &	 $0.7492$ & $0.0849$ \\
$40$	& $0.7588$ &	$0.0690$ &	 $0.7794$ & $0.0721$ \\
$50$	& $0.7568$ &	$0.0495$ &	 $0.7561$ & $0.0602$ \\
$60$	& $0.7627$ &	$0.0417$ &	 $0.7546$ & $0.0445$ \\
$70$	& $0.7558$ &	$0.0351$ &	 $0.7608$ & $0.0360$ \\
$80$	& $0.7509$ &	$0.0322$ &	 $0.7581$ & $0.0303$ \\
$90$	& $0.7498$ &	$0.0251$ &	 $0.7595$ & $0.0242$ \\
$100$& $0.7531$ &	$0.0161$ &	 $0.7586$ & $0.0167$ 
\enddata
\end{deluxetable}

\subsubsection{Fitting results}

During the fitting procedure, we have set the threshold in Eq.(\ref{equ:threpca}) to be $r_p > 99.99\%$ for the \textit{PCA-model}. For the \textit{MDS-model}, we require either $r_p^{(1)} > 99.99\%$ or  $r_p^{(2)} > 99.99\%$ satisfied, see Eq.(\ref{equ:thremds}). Then, we get two parameters $c_1$ and $c_2$ for the \textit{PCA-model}, and one parameter $\tilde c_1$ for the \textit{MDS-model}. For comparison,  these two models are fitted to observations by using both the  Levenberg-Marquardt (LM) technique and the Markov Chain Monte Carlo (MCMC) method. The current value of Hubble parameter is fixed to be $H_0 = 70.0$km/s/Mpc. 

\begin{figure}
  \epsscale{1.0}
   \plotone{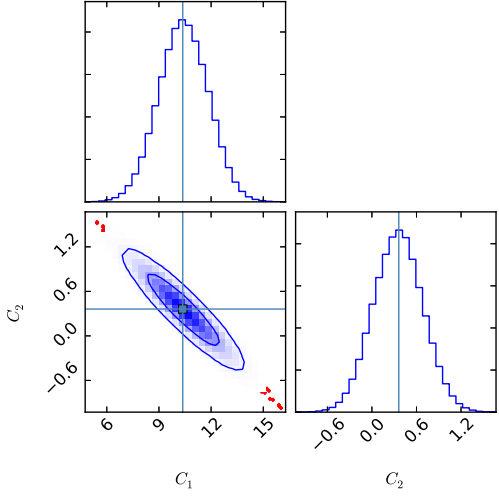}
  \caption{JLA data only. The contour from $1\sigma$ to $2\sigma$ confidence levels  and $1$-D histograms for parameters $c_1, c_2$ of the \textit{PCA-model}. The correction between $c_1$ and $c_2$ comes from the constraint $E(0) = 1$, see the definition of $E(z)$  below Eq.(\ref{equ:lumdis}).}
  \label{fig:pcatri}
\end{figure}

At first, only JLA data is used  to fit models.  After marginalizing the nuisance parameters of JLA, we obtain $c_1= 10.37\pm 1.35\,, \quad  c_2 = 0.3617 \pm  0.3157$ with  $\chi^2_{\text{min}}/ \text{d.o.f.} = 683.001/738$ (LM), while $c_1 = 10.40 ^{+1.39}_{ -1.36}, c_2 = 0.3540^{+0.3213}_{ -0.3254}$ with  $\chi^2_{\text{min}}/ \text{d.o.f.} = 683.001/738$ (MCMC) for the \textit{PCA-model}. We obtain $\tilde c_1= 2.342 \pm 0.0163$ with  $\chi^2_{\text{min}}/ \text{d.o.f.} = 683.942/739$ (LM) , while $\tilde c_1 = 2.343 ^{+0.0167}_{ -0.0163}$ with   $\chi^2_{\text{min}}/ \text{d.o.f.} = 683.942/739$ (MCMC) for the \textit{MDS-model}. The contours  for  parameters $c_1, c_2$ of the \textit{PCA-model} and their $1$-D histograms are plotted in Fig.\ref{fig:pcatri}.

\begin{figure}
  \epsscale{1.0}
   \plotone{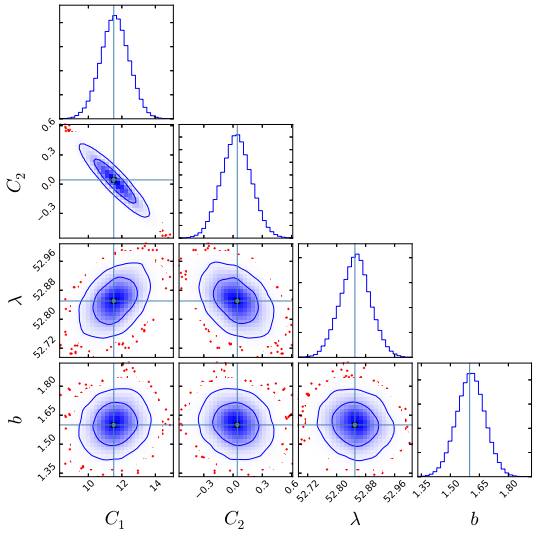}
  \caption{JLA + GRBs data . The contours from $1\sigma$ to $2\sigma$ confidence levels  and $1$-D histograms for parameters $c_1, c_2, \lambda, b$ for the \textit{PCA-model}. }
  \label{fig:pcatrigam}
\end{figure}

\begin{figure}
  \epsscale{1.2}
   \plotone{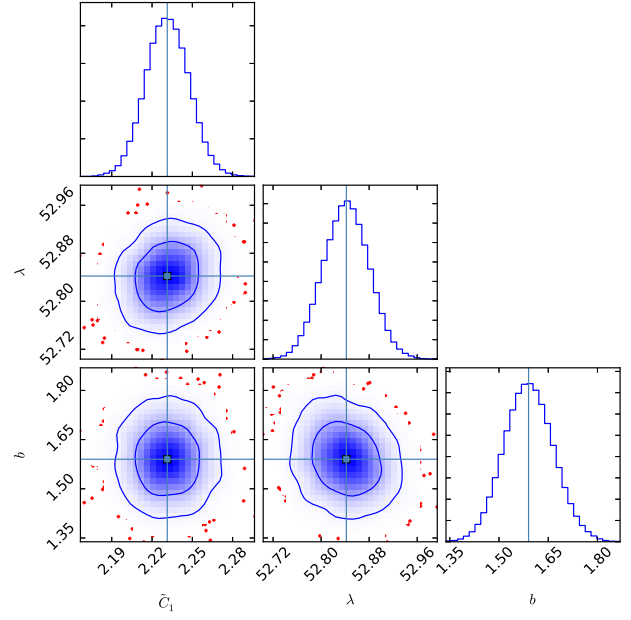}
  \caption{JLA +GRBs data. The contours from $1\sigma$ to $2\sigma$ confidence levels  and $1$-D histograms for parameters $\tilde c_1,  \lambda, b$ for the \textit{MDS-model}. }
  \label{fig:mdstrigam}
\end{figure}

Next, both JLA and GRBs data are used. The nuisance parameters of JLA is also marginalized since we do not have interest in them. However, the parameters $\lambda$ and $b$ in the Amati correlation (\ref{equ:amati}) are kept free to see how well the calibration is.  We obtain $c_1 = 11.52\pm 0.85, c_2 = 0.0434\pm 0.1546, \lambda = 52.850\pm 0.041, b = 1.600\pm 0.071$ with $\chi^2_{\text{min}} /\text{d.o.f.} = 787.592/845$ (LM), while $c_1 = 11.58^{+0.83}_{-0.82}, c_2 = 0.0329^{+0.1500}_{ -0.1519}, \lambda = 52.852^{+0.039}_{ -0.042}, b = 1.606^{+0.070}_{ -0.072} $ with $\chi^2_{\text{min}} /\text{d.o.f.} = 787.601/845$ (MCMC) for the  \textit{PCA-model}.  We obtain $\tilde c_1 = 2.231\pm 0.016, \lambda  = 52.841 \pm 0.037,  b = 1.593\pm  0.070$ with $\chi^2_{\text{min}} /\text{d.o.f.} = 787.767/846$ (LM), while $\tilde c_1 = 2.231^{+0.016}_{ -0.015}, \lambda = 52.842^{ +0.036} _{-0.038}, b = 1.590^{+0.074}_{ -0.072}$ with $\chi^2_{\text{min}} /\text{d.o.f.} = 787.769/846$ (MCMC) for the  \textit{MDS-model}. The contours  for  parameters $c_1, c_2, \lambda, b$ and their $1$-D histograms are plotted in Fig.\ref{fig:pcatrigam} for the \textit{PCA-model}, while The contours  for  parameters $\tilde c_1, \lambda, b$ and their $1$-D histograms are plotted in Fig.\ref{fig:mdstrigam} for the \textit{MDS-model}.


\begin{figure}
  \epsscale{1.2}
   \plotone{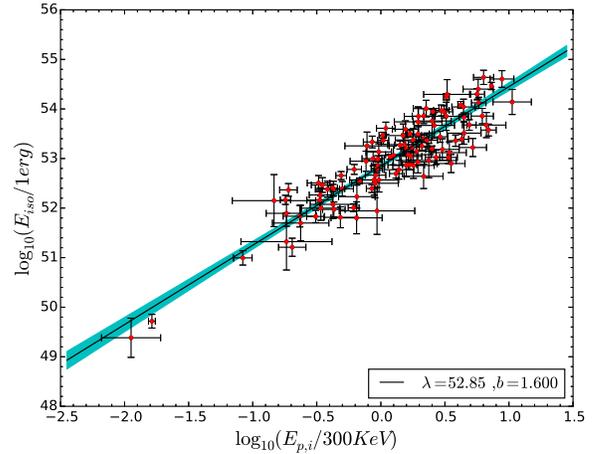}
  \caption{Calibration. The red points with error bars corresponds to $109$ GRBs data, while the line  corresponds the best-fit calibration with $1\sigma$ confidence level.}
  \label{fig:calibrate}
\end{figure}

The calibration of $109$ GRBs data is also shown in Fig.\ref{fig:calibrate}, in which the propagated uncertainties of $\log_{10} E_{\text{iso}} $ and $\log_{10} E_{\text{p,i}} $ are estimated by
\begin{eqnarray}
	\sigma_{\log_{10} E_{\text{iso}} } &=& \sqrt{\sigma_{\lambda}^2 + \sigma_b^2 \left[  \log_{10} \left( \frac{ E_{\text{p,i}} }{ \text {300KeV} } \right) \right]^2 
	+ b^2\sigma_{\log_{10} E_{\text{p,i}} }^2 } \,, \\
	\sigma_{\log_{10} E_{\text{p,i}} } &=& \frac{1}{\ln 10} \frac{\sigma_{E_{\text{p,i}}} }{E_{\text{p,i}}} \,.
\end{eqnarray}
It is clear that the calibration in this work is well consistent with data.

\subsection{Reconstruction of History}

\begin{figure}
  \epsscale{1.0}
   \plotone{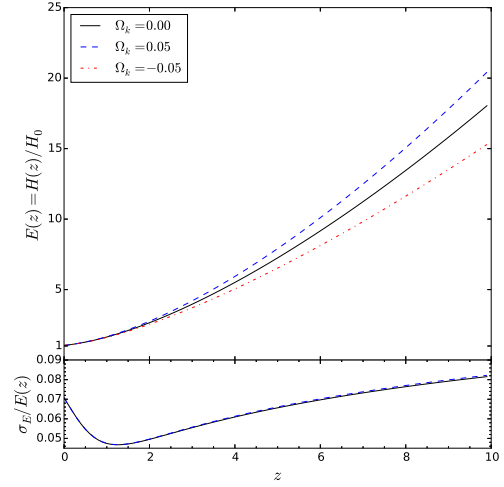}
  \caption{ Evolution history of the Universe. Top: the reconstructed cosmic expansion rate from the PCA-model. Bottom:  relative errors of $E(z)$. }
  \label{fig:pcah}
\end{figure}

Now, we are ready to reconstruct the history of Universe. The cosmic expansion rates $E(z) = H(z)/H_0$ with different spatial curvatures are plotted in Fig.\ref{fig:pcah} and Fig.\ref{fig:mdsh}.  The relative errors of $E(z)$ is estimated by
\begin{equation}\label{equ:estierr}
	\frac{\sigma_E}{E} =  \sqrt{ \left(\frac{\Omega_k D^2}{1+\Omega_k D^2}\right)^2 \left(  \frac{\sigma_{\Omega_k}^2}{4\Omega_k^2} + \frac{\sigma_D^2}{D^2}  \right)
	+ \frac{\sigma_{D'}^2}{ D^{'2}} } \,,
\end{equation}
where $\sigma_D = \sqrt{\mathbf{w}_i^2 \sigma_{c_i}^2}$,  $\sigma_{D'} = \sqrt{\mathbf{w}_i^{'2} \sigma_{c_i}^2}$ for the PCA-model, and $\sigma_D = \sqrt{\mathbf{\tilde w}_i^2 \sigma_{ \tilde c_i}^2}$,  $\sigma_{D'} = \sqrt{\mathbf{\tilde w}_i^{'2} \sigma_{\tilde c_i}^2}$ for the MDS-model . 

From Fig.\ref{fig:pcah} and Fig.\ref{fig:mdsh} , one can see that the relative error of $E(z)$ in the MDS-model is about ten times less than that in the PCA-model. And that is as it should be, because there is one parameter $\tilde c_1$ in the MDS-model, while there are two parameters $c_1$ and $c_2$ in the PCA-model. In both models, the relative error of $E(z)$ is small at low redshifts, say $0.5<z<1.0$, since most of the data points belong to this range of redshifts. 

It is interesting to see that in the MDS-model, the relative error of $E(z)$ is a constant for a spatial-flat Universe ($\Omega_k = 0$). This could be seen from Eq.(\ref{equ:estierr}): $\sigma_E/E = \sigma_{D'}/ D' = \sigma_{\tilde c_1}/\tilde c_1$, since there is only one parameter $\tilde c_1$ in the MDS-model.  Taking the best fitting value for $\tilde c_1$ and its uncertainty $\sigma_{\tilde c_1} $, we obtain $\sigma_E/E \approx 0.7\%$. 

In Fig.\ref{fig:pcah} and Fig.\ref{fig:mdsh}, the spatial curvature $\Omega_k$ is chosen to show the differences of the cosmic expansion rate under different space geometries. In fact, \citet{Li2014} have already taken a model-independent approach to determine the spatial curvature  by using the recent baryon acoustic oscillation (BAO) measurements. According to their conclusions,  the errors of $\Omega_k$ decrease with increasing redshift, and the best constraint is $\Omega_k = -0.05 \pm 0.06$ (at $z=2.36$ ).  However the errors of curvature at low redshifts are nearly of order unit, see \citet[Fig.2]{Li2014} .  Considering the future BAO measurements, at least one order of magnitude improvement of $\Omega_k$ could be expected at both low and high redshifts \citep{Li2014}. 

\begin{figure}
  \epsscale{1.0}
   \plotone{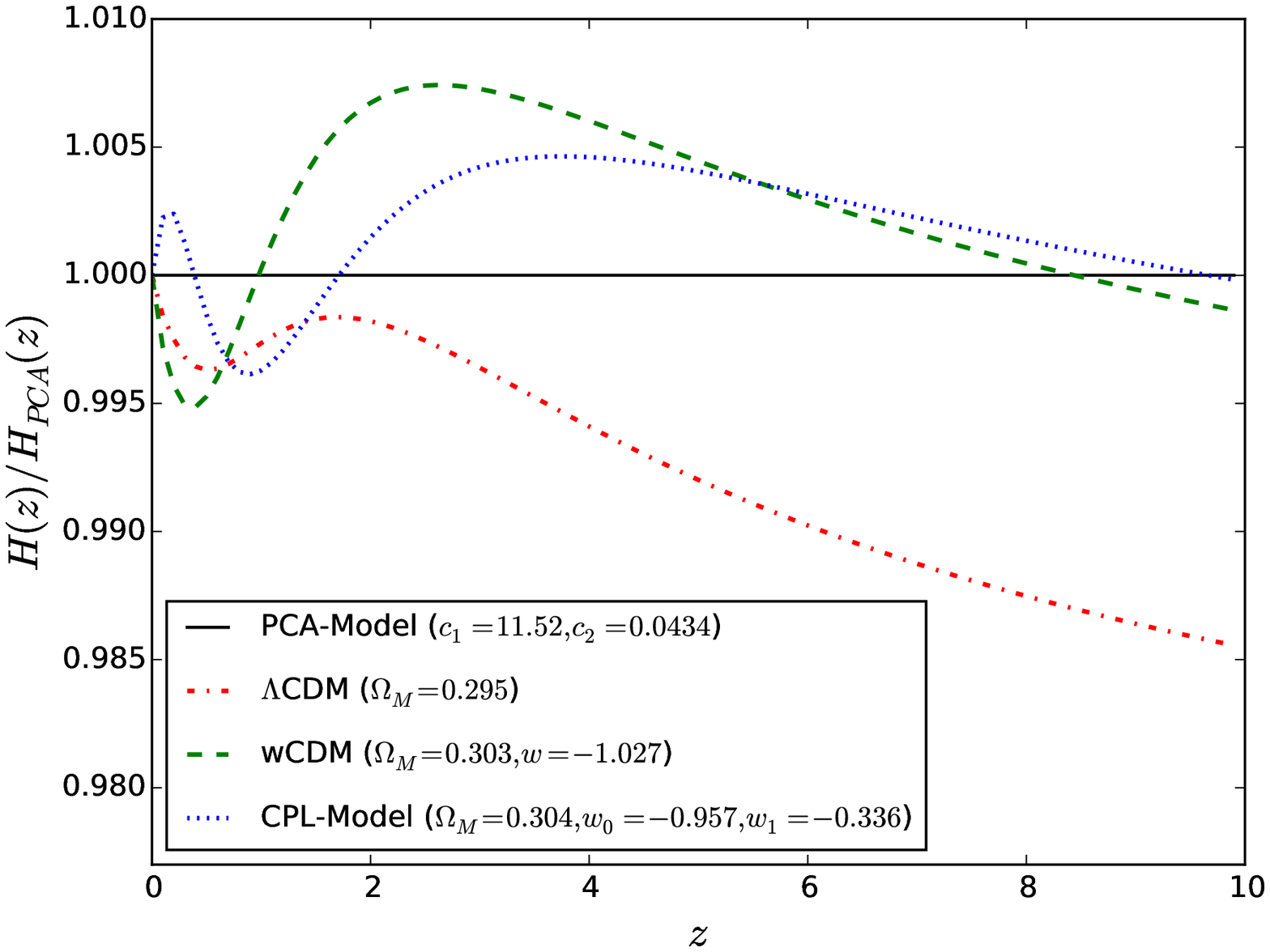}
  \caption{ Comparison of the cosmic expansion rate predicted from the $\Lambda$CDM, the wCDM and  the CPL model with their best fitting parameters in Ref. \citep{Benitez2013} and that reconstructed from the PCA-model, namely, $H(z)/H_{PCA}(z)$. }
  \label{fig:pcahc}
\end{figure}

\begin{figure}
  \epsscale{1.0}
   \plotone{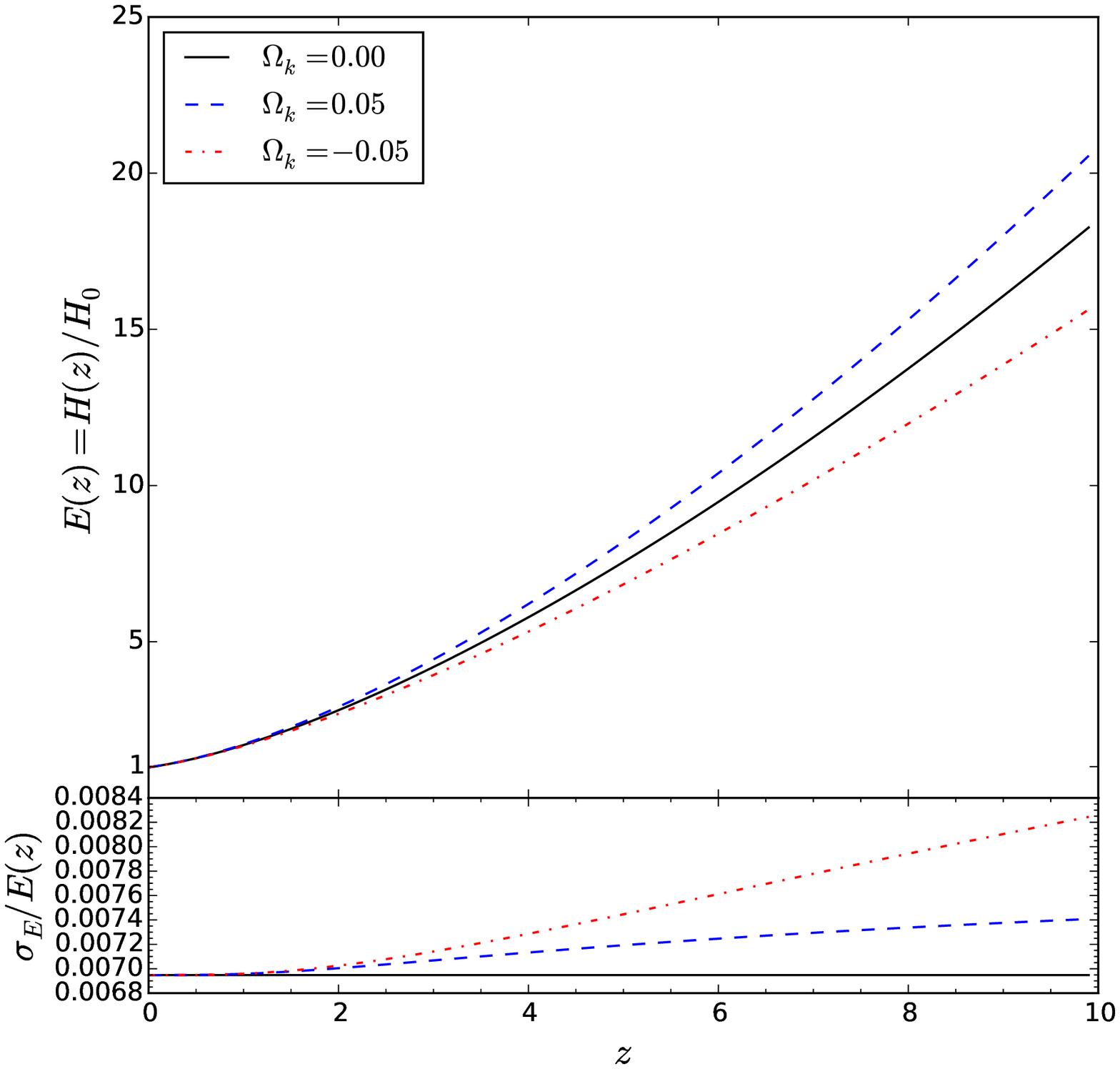}
  \caption{ Evolution history of the Universe. Top: the reconstructed cosmic expansion rate from the MDS-model. Bottom:  relative errors of $E(z)$. }
  \label{fig:mdsh}
\end{figure}

\begin{figure}
  \epsscale{1.0}
   \plotone{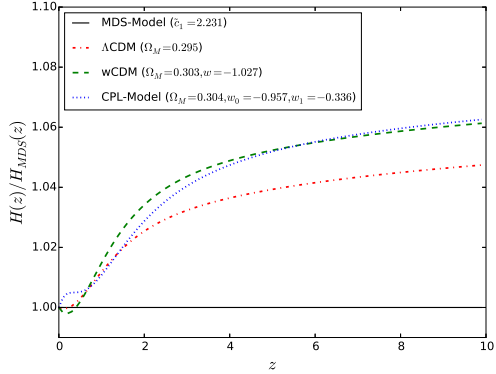}
  \caption{ Comparison of the cosmic expansion rate predicted from the $\Lambda$CDM, the wCDM and  the CPL model with their best fitting parameters in Ref. \citep{Benitez2013} and that reconstructed from the MDS-model, namely, $H(z)/H_{MDS}(z)$. }
  \label{fig:mdshc}
\end{figure}

The ratio of the cosmic expansion rate that predicted from the $\Lambda$CDM, the wCDM and  the CPL model with their best fitting parameters in Ref. \citep{Benitez2013} to that reconstructed from the PCA-model and the MDS-model, i.e. $H(z)/H_{PCA}(z)$ and $H(z)/H_{MDS}(z)$ are plotted in  Fig.\ref{fig:pcahc} and Fig.\ref{fig:mdshc} respectively. 

From Fig.\ref{fig:pcahc}, one can see that  in the $\Lambda$CDM model the expansion rate is always smaller than that in the PCA-model, while in the wCDM model $H(z)$ is firstly smaller than $H_{PCA}(z)$ at low redshifts,  then  becomes  larger than $H_{PCA}(z)$  at medium redshifts, and finally gets smaller than $H_{PCA}(z)$ again at high redshifts. In the CPL model, the behavior of $H(z)$ is almost like that in the wCDM model except that $H(z)$ is firstly larger than $H_{PCA}(z)$. From Fig.\ref{fig:mdshc}, one can see that the behavior of $H(z)$ in these three physical models are almost the same as each other except a small difference at very low redshifts, and they are larger than $H_{MDS}(z)$ at a large range of the redshifts. 

Due to the precision limit, we can not find out these differences discussed above from the present observations, since these differences are really quite small.

\section{Discussions and Conclusions}\label{sec:discon}. 

Cosmological variables such as the luminosity distance can be decomposed into some suitable basis. In this paper, we have proposed two methods: PCA and MDS to optimize this basis. The projected feature spaces that describe the luminosity distance could then retain most of the origin information in a low-dimensional space. We call them the \textit{PCA-model} and the \textit{MDS-model} respectively. It should be noticed that the procedures used above do not depend on any specific cosmological models. After that, 
observational data including the "joint light-curve analysis" (JLA) data set that consists of 740 Type Ia supernovae (SNIa) as well as 109 long gamma-ray bursts with the well-known Amati relation  are used to constrain the parameters of these two models by using the Levenberg-Marquardt  technique and the Markov Chain Monte Carlo  method . Finally, we obtain the evolution history of the Universe including both the cosmic expansion rate and its relative errors and we also compare the results with that  predicted from the $\Lambda$CDM, the wCDM and  the CPL model with their best fitting parameters.

We notice that whether the  \textit{PCA-model} or the  \textit{MDS-model} could be used to perform the calibration to GRBs data without any prior assumptions of a specific cosmological model. We also estimate the system errors of GRBs data.  We can say with confidence that the error bars will become smaller when more accurate GRBs data would be obtained in the future.


\acknowledgments
CJF would like to thank Puxun Wu for helpful discussions. This work is supported by National Science Foundation of China grant Nos.~11105091 and~11047138, ``Chen Guang" project supported by Shanghai Municipal Education Commission and Shanghai Education Development Foundation Grant No. 12CG51, National Education Foundation of China grant  No.~2009312711004, Shanghai Natural Science Foundation, China grant No.~10ZR1422000, Key Project of Chinese Ministry of Education grant, No.~211059,  and  Shanghai Special Education Foundation, No.~ssd10004, and the Program of Shanghai Normal University.





\clearpage

\begin{turnpage}
 \begin{deluxetable}{llllllll}
\tabletypesize{\scriptsize}
\tablecaption{The efficiency of PCA and MDS for different TSGs\label{tab:eff}}
\tablewidth{0pt}
\tablehead{
\colhead{TSGs} & \multicolumn{2}{c}{PCA} &  \multicolumn{4}{c}{MDS} & \colhead{Parameters} \\
\colhead{}
& \colhead{$\langle r_1 \rangle$}           & \colhead{$\langle r_2 \rangle$}
& \colhead{$ \langle r_1^{(1)} \rangle$ } & \colhead{ $ \langle r_2^{(1)} \rangle$ }
& \colhead{ $\langle r_1^{(2)} \rangle$ } & \colhead{ $ \langle r_2^{(2)} \rangle$ }
& \colhead{}
}
\startdata
 $\Lambda$CDM   
 & $99.834\%$ & $99.996\%$ & $99.692\%$ & $99.998\%$ &$99.998\%$& $99.999\%$    & $0.1<\Omega_M<0.9$, $0.1<\Omega_V<0.9$, $-0.1<\Omega_K<0.1$\\
 wCDM 		     
 & $99.859\%$ & $99.995\%$ & $99.413\%$ & $99.992\%$ &$99.995\%$& $99.999\%$    & $0.1<\Omega_M<0.9$, $0.1<\Omega_V<0.9$, $-0.1<\Omega_K<0.1$, $-1.5<w<-0.5$\\
 CPL    \tablenotemark{a}		     
 & $99.906\%$ & $99.998\%$ & $99.701\%$ & $99.997\%$ &$99.998\%$& $99.999\%$& $0.1<\Omega_M<0.9$, $0.1<\Omega_V<0.9$, $-1.5<w_0<-0.5$, $-0.5<w_1<0.5$\\
 FSLL-I  \tablenotemark{b}		     
 & $99.895\%$ & $99.997\%$ & $99.767\%$ & $99.997\%$ &$99.998\%$& $99.999\%$& $0.1<\Omega_M<0.9$, $0.1<\Omega_V<0.9$, $-1.5<w_0<-0.5$, $-0.5<w_1<0.5$\\
 FSLL-II \tablenotemark{b}			     
 & $99.905\%$ & $99.998\%$ & $99.795\%$ & $99.998\%$ &$99.998\%$& $99.999\%$& $0.1<\Omega_M<0.9$, $0.1<\Omega_V<0.9$,  $-1.5<w_0<-0.5$, $-0.5<w_1<0.5$\\
 HDE   \tablenotemark{c} 		     
 & $99.922\%$ & $99.999\%$ & $99.460\%$ & $99.996\%$ &$99.997\%$& $99.999\%$& $0.1<\Omega_M<0.9$, $0.1<\Omega_V<0.9$, $0.1<C<1.5$\\
 DGP   \tablenotemark{d}		     
 & $99.804\%$ & $99.999\%$ & $99.260\%$ & $99.998\%$ &$99.992\%$& $99.999\%$& $0.1<\Omega_M<0.9$, $0.1<\Omega_V<0.9$\\
 NADE  \tablenotemark{e}		     
 & $99.897\%$ & $99.999\%$ & $99.706\%$ & $99.998\%$ &$99.998\%$& $99.999\%$& $0.1<\Omega_M<0.9$, $0.1<\Omega_V<0.9$, $1.5<n<3.5$\\
 RDE    \tablenotemark{f}		     
 & $99.947\%$ & $99.998\%$ & $99.848\%$ & $99.995\%$ &$99.998\%$& $99.999\%$& $0.1<\Omega_M<0.9$, $0.1<\Omega_V<0.9$, $0.1<\alpha<1.0$\\
 Mixture		     
 & $99.780\%$ & $99.995\%$ & $99.250\%$ & $99.998\%$ &$99.995\%$& $99.999\%$& Take the same ranges as above.
\enddata
\tablenotetext{a}{\citet{Chevallier2001, Linder2003}}
\tablenotetext{b}{\citet{Feng2012}}
\tablenotetext{c}{\citet{Li2004}}
\tablenotetext{d}{\citet{Dvali2000, Deffayet2001,Deffayet2002} }
\tablenotetext{e}{\citet{Wei2008} }
\tablenotetext{f}{\citet{Gao2009, Feng2009} }
\end{deluxetable}
\end{turnpage}
\clearpage

\end{document}